\documentclass[amstex,twocolumn,floats,floatfix,aps,pra, nofootinbib]{revtex4-2}
\usepackage{amssymb}
\usepackage{amsmath}
\usepackage{calc}
 \usepackage{graphicx}
\usepackage{bm}
\usepackage{appendix}
\usepackage{titlesec}
\usepackage{soul}
\usepackage{xcolor}
\usepackage{comment}

\sethlcolor{yellow}

\titleformat{\paragraph}
{\normalfont\normalsize}{\theparagraph}{1em}{}
\titlespacing*{\paragraph}
{0pt}{3.25ex plus 1ex minus .2ex}{1.5ex plus .2ex}

\def\be{ \begin{equation} }
\def\ee{ \end{equation} }
\def\bf{ \begin{figure} }
\def\ef{ \end{figure} }
\def\bea{ \begin{eqnarray} }
\def\eea{ \end{eqnarray} }
\def\bse{ \begin{subequations} }
\def\ese{ \end{subequations} }
\def\sech{\,\text{sech}}

\def\d{\,\text{d}}

\def\P01{ P_{\text{0}\rightarrow \text{1}} }

\def\ket#1{\vert #1\rangle}

\begin{document}

\author{Ivo S. Mihov}
\affiliation{Department of Physics, Sofia University, James Bourchier 5 blvd, 1164 Sofia, Bulgaria}
\author{Nikolay V. Vitanov}
\affiliation{Department of Physics, Sofia University, James Bourchier 5 blvd, 1164 Sofia, Bulgaria}
\title{Pulse shape effects in qubit dynamics demonstrated on an IBM quantum computer}
\date{\today }

\begin{abstract}
We present a study of the coherent interaction of a qubit with a pulse-shaped external field of a constant carrier frequency. 
We explore, theoretically and experimentally, the transition line profile --- the dependence of the transition probability on the detuning --- for five different pulse shapes: rectangular, Gaussian, hyperbolic-secant, squared hyperbolic-secant and exponential. 
The theoretical description for all cases but sech$^2$ is based on the analytical solutions to the Schr\"odinger equation or accurate approximations available in the literature. 
For the sech$^2$ pulse we derive an analytical expression for the transition probability using the Rosen-Zener conjecture, which proves very accurate. The same conjecture turns out to provide a very accurate approximation for the Gaussian and exponential pulses too.
The experimental results are obtained with one of IBMQ's quantum processors. 
An excellent agreement between theory and experiment is observed, demonstrating some pulse-shape-dependent fine features of the transition probability profile. 
The mean absolute error --- a measure of the accuracy of the fit --- features an improvement by a factor of 4 to 8 for the analytic models compared to the commonly used Lorentzian fits.
Moreover, the uncertainty of the qubit's resonance frequency is reduced by a factor of 4 for the analytic models compared to the Lorentzian fits. 
These results demonstrate both the accuracy of the analytic modelling of quantum dynamics and the excellent coherent properties of IBMQ's qubit.
\end{abstract}

%\pacs{03.65.Ge, 32.80.Bx, 34.70.+e, 42.50.Vk}
\maketitle

%%%%%%%%%%%%%%%%%%%%%%%%%%%%%%%%%%%%%%%%%%%%%%%%%%%%%%%%%%%%%%%%%%%%%%%%%%%%%%%%%%%%%%%%%%%%%%%
%%%%%%%%%%%%%%%%%%%%%%%%%%%%%%%%%%%%%%%%%%%%%%%%%%%%%%%%%%%%%%%%%%%%%%%%%%%%%%%%%%%%%%%%%%%%%%%
%%%%%%%%%%%%%%%%%%%%%%%%%%%%%%%%%%%%%%%%%%%%%%%%%%%%%%%%%%%%%%%%%%%%%%%%%%%%%%%%%%%%%%%%%%%%%%%
\section{Introduction}

Quantum algorithms promise to revolutionise the way we do certain operations like searching, counting, factoring, encrypting and transferring data, and even breaking some advanced encryption protocols. 
Quantum algorithms are constructed using quantum gates, which in turn are achieved by applying a pulse of microwave or laser radiation to the qubit. 
The behaviour of the qubit under different types of pulses is described by a number of models, each defined by the time dependence of the Rabi frequency $\Omega(t)$ and the detuning $\Delta(t)$ of the applied pulse. 
One common trait of all models is they predict the pulse area $S=\int_0^T\Omega(t)\,dt$ must be equal to an odd multiple of $\pi$ for unit transition probability to state $\ket{1}$, assuming we start in state $\ket{0}$ and we are exactly at resonance ($\Delta=0$).

In general, we can categorise the models in several ways. 
One option is to distinguish ones with constant or variable detuning. 
From another viewpoint, we can divide them into 3 different categories depending on the pulse shape: exponential, trigonometric and polynomial. 
Among the exponential models, we have the Rosen-Zener model~\cite{Rosen1932}, the Gaussian model~\cite{Vasilev2004}, the Tanh model~\cite{Simeonov2014}, and the Demkov model~\cite{Demkov1963}. 
Among the trigonometric ones we have the powers-of-sine models~\cite{Boradjiev2013} (e.g. the Sine model, the Sine-squared model etc.). 
The polynomial models include the Lorentzian-based models~\cite{Vasilev2014}. 
Note that the Rabi (rectangular) model~\cite{Rabi1937} can be treated as a zeroth-order member of any of these categories, so it forms its own group.

Certain types of models assume pulses of infinite duration. 
Naturally, they are not exactly applicable in practice. 
However, they can be modified, so as to adopt a truncated version of the same type of pulse, leading to practical models with actual implications. 
Examples of such pulse types include the hyperbolic secant, Gaussian, Lorentzian 
and other infinite-duration shapes. 
Other models have natural initial and final points and do not need truncation. 
This usually implies smaller dependence on the actual implementation of the model. 
Among these are the Rabi model, the Sine model, etc. 

The pulse shape plays a key role in the functional form of the transition probability. 
Specifically, this affects the transition line profile $\P01(\Delta)$ --- the 
dependence of the transition probability $\P01$ on the detuning $\Delta$. 

Although pulse shape effects on the qubit dynamics have been known for a long time, very few studies have directly verified them experimentally.
Among these we note the significant reduction of power broadening in two-photon excitation by Gaussian pulses~\cite{Vitanov2001,Halfmann2003} and a detailed study of pulse-shape effects on the Rabi oscillations~\cite{Conover2011}.

We note that on exact resonance, the transition probability depends on the pulse area 
only and hence the ensuing Rabi oscillations do not depend on the pulse shape.
It is off-resonance where the difference begins to emerge --- the larger the detuning 
from resonance, the greater the differences.
Hence, the most direct way to observe pulse-shape effects is to explore the transition 
line profile. % - the dependence of the transition probability of the detuning.
This task is somewhat obstructed by the fact that for large detuning (much greater than 
$1/\tau$, the inverse pulse width) where the pulse-shape differences are most pronounced, 
the transition probability is very small and hard to measure. 
Nevertheless, measurable differences emerge even for small detuning (of the order of $1/\tau$).

In the last few years, a new trend in quantum computing has emerged. With the 
appearance of remotely accessible quantum computing devices like IBM Quantum 
Experience~\cite{IBM}, it became possible to perform quantum control 
demonstrations with them \cite{Torosov2022,Ivanov2022}. Moreover, other influential 
quantum computing demonstrations on IBM Quantum have recently been conducted in other quantum 
computing fields \cite{Skosana2021, Pokharel2022, Pivoluska2022, Ezzell2022, Tripathi2021}. 

In this work, we explore the notion that the pulse shape has a pronounced impact 
on the transition line profile. 
We study the transition probability for several exponential models of constant detuning, 
for which exact or approximate expressions for the transition probability are available 
--- the Rabi, Rosen-Zener, Gaussian, and Demkov models. 
We add to these the sech$^2$ model, for which the transition probability is derived 
here by using an earlier theoretical conjecture. 
We compare the theoretical predictions with experimental results obtained by us using 
the ibm\_perth quantum processor~\cite{IBM}. 
Evaluating the mean absolute error~(MAE), we give a measure of the extent to which 
the theoretical models apply to the experimental measurements.

%%%%%%%%%%%%%%%%%%%%%%%%%%%%%%%%%%%%%%%%%%%%%%%%%%%%%%%%%%%%%%%%%%%%%%%%%%%%%%%%%%%%%%%
\section{Experimental demonstration}
\label{sect-method}

In this work we use qubit 0 of ibm\_perth, one of the IBM Quantum Falcon r5.11H
Processors~\cite{IBM}. It is an open-access quantum processor that consists of seven transmon qubits~\cite{Koch2007} 
The exact date of the demonstration is 26 May 2023. At the time of the 
demonstration the parameters of the zeroeth qubit of the ibm\_perth system are 
calibrated as follows: the qubit frequency is $5.15754$~GHz, with anharmonicity 
$-0.34152$~GHz. The T1 coherence time is  $166.70~\mathrm{\mu s}$, the T2 
coherence time is $91.32~\mathrm{\mu s}$ and the readout assignment error 
is $2.65$\%.

For the purpose of the manuscript we evaluate the measured frequency responses for five different pulse shapes. 
All the pulse shapes are centred at $t = T / 2$, where $T$ is the duration of the pulse. Shape symmetry implies that the pulses have truncation points at $t=0$ and $t=T$. In order to minimise the impact of truncation on the results, the duration $T$ is picked so that the Rabi frequency value at the truncation points is about 0.1\% of the maximum Rabi frequency $\Omega_0$ of the corresponding pulse. This way, we push the truncation far from the centre of the pulse and limit the magnitude of the discontinuous jump to $\Omega_0/1000$.

Each of the transition line profiles is then fitted with an analytically calculated expression and also with a Lorentzian function for comparison. 
The Lorentzian fit was inspired from Ref.~\cite{ibm-textbook} where it is used to find the resonant frequency by fitting transition line profiles of Gaussian pulses in order to subsequently calibrate X gates.
The analytical expressions are calculated using the selected values for the pulse width, duration and Rabi frequency of the excitation pulse, and then are post-processed to take into account the effects of the dephasing, readout and leakage errors of the qubit, see the text below. 
Furthermore, the fits are checked for overfitting by calculating an overfitting index using an average of the point-wise derivative of the measurements.

%%%%%%%%%%%%%%%%%%%%%%%%%%%%%%%%%%%%%%%%%%%%%%%%%%%%%%%%%%%%%%%%%%%%%%%%%%%%%%%%%%%%%%%
\subsection{Dephasing, readout and leakage errors} \label{sect-postprocess}

The post-processing procedure takes three effects --- dephasing, readout error and population leakage to non-qubit states ---  into account in an empirical manner.
Because these losses affect the qubit states 0 and 1 differently, we define the transition probability as 
\be
    \label{eq-readout}
    \P01 = \epsilon_0 + (1-\epsilon_0-\epsilon_1)\P01^{(0)}
\ee
where $\P01^{(0)}$ is the ideal transition probability in the completely coherent, lossless case, whereas $\epsilon_0$ and $\epsilon_1$ are free fitting parameters.
Obviously, if $\P01^{(0)} = 0$, we have $\P01 = \epsilon_0$, and if $\P01^{(0)} = 1$, we have $\P01 = 1-\epsilon_1$, i.e., $\epsilon_0$ accounts for the lifted background and $\epsilon_1$ for the degradation of the probability from unity.

The modification of the transition probability of Eq.~\eqref{eq-readout} was always applied, with $\P01^{(0)}$ taken from the theoretical expressions for the transition probability with no fitting parameters, and  $\epsilon_0$ and $\epsilon_1$ determined by fitting to the transition probability values.

%%%%%%%%%%%%%%%%%%%%%%%%%%%%%%%%%%%%%%%%%%%%%%%%%%%%%%%%%%%%%%%%%%%%%%%%%%%%%%%%%%%%%%%%%%%%%%%
\subsection{Fits and agreement between theory and observations: divergence and overfitting indices}
\label{sect-fits}

The measured data points $P(\Delta) $ for each of the models executed in the demonstration are fit to two different shapes --- a model formula $\P01(\Delta)$ for each pulse shape and a Lorentzian $L(\Delta)$. 
The model prediction is then compared to the Lorentzian by finding the residuals $R(\Delta)$ for some fit $F(\Delta)$ as
\be    \label{eq-residuals}
    R(\Delta) = F(\Delta) - P(\Delta),
\ee
where $F(\Delta)$ can be either the model $\P01(\Delta)$ or the Lorentzian fit $L(\Delta)$.

Using Eq.~\eqref{eq-residuals} we calculate the mean absolute error~(MAE)
\be
    \label{eq-divergence}
    \text{MAE} = \frac{1}{N}\sum_{i=1}^N |R(\Delta_i)| = \frac{1}{N} \sum_{i=1}^N %
    |F(\Delta_i) - P(\Delta_i)|
\ee
where a larger MAE signals greater discrepancy between the fit and the observations. 

Sometimes fitting a function with many parameters to a small number of data points can lead to overfitting. 
This is a phenomenon where the fit describes the training data very well, but does not reproduce the behaviour well, as new data points that are not in the training set often have large deviations. 

To check whether we have overfitting we find the discrete derivative of the transition probability measured values and take the absolute value, then take the mean to find the overfitting index (OFI), viz.
\be
    \label{eq-of}
    \text{OFI} = \frac{1}{N-1} \sum_{k=1}^{N-1} |P(\Delta_{k+1}) - %
    P(\Delta_{k})|.
\ee
We compare the OFI against a threshold value to determine if we have overfitting. 
The overfitting threshold value is empirically chosen to be $0.1$.
If the OFI is larger than the threshold, we identify overfitting and repeat the fitting process with new initial parameters given to the fit.

%%%%%%%%%%%%%%%%%%%%%%%%%%%%%%%%%%%%%%%%%%%%%%%%%%%%%%%%%%%%%%%%%%%%%%%%%%%%%%%%%%%%%%%
\subsection{Different pulse envelopes}%
\label{sect-pulse_shapes}
Multiple pulse shapes are investigated in this work. 
Among them are three examples of exactly analytically soluble models --- the Rabi~\cite{Rabi1937}, Rosen-Zener~\cite{Rosen1932} and Demkov~\cite{Demkov1963} models --- and two models that only have approximate solutions: the Gaussian~\cite{Vasilev2004} and sech$^2$ models. 

The models described in this work have been implemented with one of IBM's quantum processors, ibm\_perth~\cite{IBM}, as described in Sec.~\ref{sect-method}. 
All the instructions were compiled and sent to the quantum processor using IBM's Qiskit Pulse package~\cite{Aleksandrowicz2019}. 
This presented an opportunity for full control of the microwave pulses applied to the transmon qubit. 
An important detail of the Qiskit Pulse framework is that it demands that the pulses are not fully continuous, but discretised into small time units of duration 2/9~ns. 
Each transition probability measurement was repeated 4096~times to reduce the impact of random deviations on the final data points. 
The pulses were chosen to have the same pulse width to be able to compare the effects between their transition line profiles directly. Furthermore, their durations were chosen such that the Rabi frequency values at the point of truncation are 0.1\% of the maximum value.
All the pulses used in this work have a temporal pulse area of $\pi$, i.e., they are supposed to produce complete population transfer on exact resonance in the absence of decoherence and other losses.

The data was fitted with two or three different expressions: one or two analytic formulae for the transition probability %from Sect.~\ref{sect-pulse_shapes}
and a Lorentzian $L = A\,/\,(1 + (k\Delta^2)) + C$ that was fit to each measured transition line profile.
The Lorentzian was inspired from previously used functions in the literature~\cite{ibm-textbook}.

%%%%%%%%%%%%%%%%%%%%%%%%%%%%%%%%%%%%%%%%%%%%%%%%%%%%%%%%%%%%%%%%%%%%%%%%%%%%%%%%%%%%%%%
\section{Pulse shape effects on qubit transition profile}\label{sect-bg}

The governing equation for a lossless two-state system is the Schr\"{o}dinger equation~\cite{Shore1990} with probability amplitudes $\mathbf{c}(t) = \left[c_1(t), c_2(t)\right]^T$ 
\begin{equation}
    i\hbar\frac{d}{dt}\mathbf{c}(t)=\mathbf{H}(t)\mathbf{c}(t),
    \label{eq-schr}
\end{equation}
where $\mathbf{H}(t)$ is the Hamiltonian, in the rotating-wave approximation (RWA),
\begin{equation}
    \mathbf{H}(t) = \hbar \left[ \begin{array}{cc}
    0 & \frac{1}{2}\Omega (t)\\
    \frac{1}{2}\Omega (t) & \Delta (t)
    \end{array}\right] ,
    \label{eq-ham}
\end{equation}
where $\Delta(t)$ is  the detuning and $\Omega(t)$ is the Rabi frequency. 
Eq.~\eqref{eq-ham} allows to find exact analytic solutions for different 
pairs of time-dependent $\Omega(t)$ and $\Delta(t)$. 
Of these, we select the three available models with a constant detuning, $\Delta(t)=$\,const. 
Each of the three models involves a specific time-dependent Rabi frequency, i.e. a 
specific pulse shape: rectangular (Rabi model), hyperbolic-secant (Rosen-Zener model) 
and exponential (Demkov model). 
We also use two other pulse shapes, which do not allow for exact analytic solutions but only approximate ones: Gaussian and squared hyperbolic secant.

%%%%%%%%%%%%%%%%%%%%%%%%%%%%%%%%%%%%%%%%%%%%%%%%%%%%%%%%%%%%%%%%%%%%%%%%%%%%%%%%%%%%%%%
\subsection{Rectangular pulse: Rabi model}

\subsubsection{Transition probability}
In the Rabi model~\cite{Rabi1937}, the Rabi frequency takes a constant form within 
a finite time interval and is zero outside, 
\begin{equation}    
    \label{eq-rabi-def}
        \Omega(t) = \left\{ 
        \begin{array}{cc}
            \Omega_0, & \text{for } 0\leq t \le T;\\
            0, & \text{elsewhere,} 
        \end{array}\right.
\end{equation}
where $T$ is the duration of the pulse, and the detuning $\Delta$ is constant. 

The transition probability reads 
\be 
    \P01^{(0)}(\Delta)= 
    {\frac{\Omega_0^2}{\Delta^2+\Omega_0^2}}
    \sin^2{\left(\frac{T}{2}\sqrt{\Delta^2+\Omega_0^2}\right)}.
    \label{eq-rabi}
\ee
When plotted versus the detuning the transition probability features a central 
maximum at $\Omega_0 T=\pi$ and damped oscillations. These oscillations, or 
satellite peaks, which are characteristic for the Rabi model, result from the discontinuities at the initial and final instants of the pulse. Prior to its use as a fitting function, the transition line profile formula is modified according to Eq.~\eqref{eq-readout} to account for population losses.

\begin{figure}[tb]
    \centering
    \includegraphics[width=\linewidth]%
    {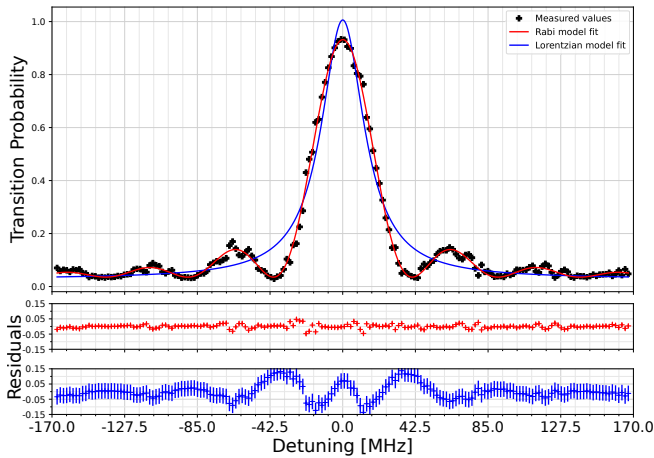}
    \caption{(Colour online) The transition line profile for the Rabi model: 
    measured data (black crosses), fit using the Rabi formula in Eq.~\eqref{eq-rabi} (red) and a Lorentzian $L = A/(1 + (k\Delta^2)) + C$
    curve (blue) for comparison, where $k$ is a free fitting parameter.}
    \label{fig-rabi}
\end{figure}

\subsubsection{Implementation}

The implementation of the Rabi model featured a rectangular pulse applied to 
the qubit over a time period $T\approx 21.33$~ns. Since the pulse area must 
be $\pi$, this dictated the choice of the Rabi frequency: $\Omega_0 \approx %
23.44$~MHz.
The corresponding amplitude value in Qiskit was $0.10083$. 
This is the only pulse shape that remains unaffected by the discretisation demanded by Qiskit Pulse, and by the initial and final cut-offs effects, applied to the other pulses.

The experimental data and the analytical fits are shown in Fig.~\ref{fig-rabi}. 
The peaks are not captured by the Lorentzian pulse.
In comparison, they are represented very well by the analytical solution for 
the transition probability of Eq.~\eqref{eq-rabi}. A plot of the residuals 
between the analytic (in red) and the Lorentzian (in blue) fits and the experimental
data are shown in the bottom of the figure. 
It is evident from the residuals plot that there are some random fluctuations in the data, which add to the error of the fits.
The mean absolute errors, obtained by 
finding the sum of the absolute values of the residuals of the fits, are $9.41 \times 10^{-3}$ 
for the analytical fit and $36.21 \times 10^{-3}$ for the Lorentzian, or roughly a four-fold improvement. 
Both fits passed the overfitting criterion successfully as the OF parameters were 
smaller than our empirically chosen limit by a factor of more than 100.
The standard deviation of the resonant frequency of the analytical fit was 
$79.8$~kHz compared to $315.2$~kHz for the Lorentzian, 
which is a four-fold decrease of the uncertainty of the resonant frequency.
The values of $\epsilon_0$ and $\epsilon_1$ for the Rabi model fit are respectively $0.036$ and $0.895$.

%%%%%%%%%%%%%%%%%%%%%%%%%%%%%%%%%%%%%%%%%%%%%%%%%%%%%%%%%%%%%%%%%%%%%%%%%%%%%%%%%%%%%%%
\subsection{Hyperbolic-secant pulse: Rosen-Zener model}
\subsubsection{Transition probability}

In the Rosen-Zener model~\cite{Rosen1932} the Rabi frequency is a hyperbolic-secant function of time,
\begin{equation}
    \label{eq-rz-def}
    \Omega(t) = \Omega_0 \sech{\left(\frac{t - {T}/{2}}{\tau}\right)} ,
\end{equation}
where $\tau > 0$ is the pulse width. The detuning $\Delta$ is constant.
The transition probability is given by the exact formula~\cite{Rosen1932}
\begin{equation}
    \label{eq-rz}
    \P01^{(0)}(\Delta) =
    \frac{\sin^2{\left(\frac{1}{2}\pi\Omega_0 \tau\right)}}{\cosh^2{\left(\frac{1}{2}\pi\Delta \tau\right)}}.
\end{equation}
A unique feature of the Rosen-Zener model is the factorisation of the dependencies on the peak Rabi frequency $\Omega_0$ and the detuning $\Delta$, a property not found in any other model.
The transition probability is corrected for experimental deviations (see Eq.~\eqref{eq-readout}) before being used for fitting the observations.

\subsubsection{Implementation}
\label{sect-exp-rz}
\begin{figure}[tb]
    \includegraphics[width=\linewidth]%
    {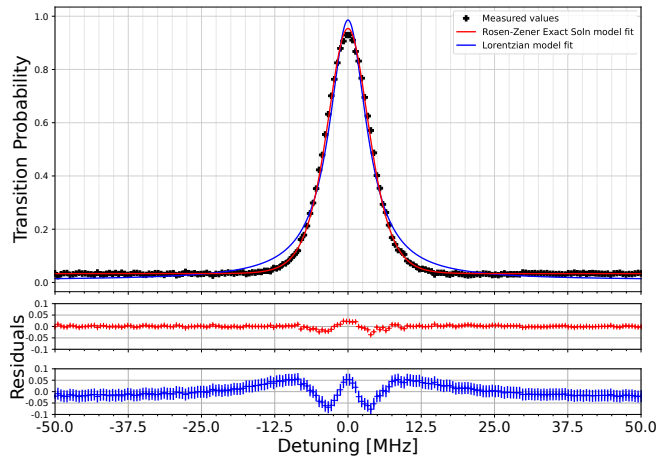}
    \caption{(Colour online) The transition line profile for the Rosen-Zener model: 
    measured data (black crosses), fit using the 
    Rosen-Zener formula in Eq.~\eqref{eq-rz} (red) and a Lorentzian 
    curve (blue) for comparison.}
    \label{fig-rz}
\end{figure}%

Contrary to the Rabi model, the Rosen-Zener model assumes a pulse of infinite duration.
Therefore, {we had to truncate the}  hyperbolic-secant pulse in the experiment. 
Moreover, unlike the Rabi model, in the Rosen-Zener model the discretisation may affect the pulse shape and its transition line profile. The pulse that was used had pulse duration (truncation point at) $T\approx324.30$~ns and pulse width $\tau\approx21.33$~ns, leading to $\Omega_0\approx 7.49$~MHz. 
This constituted a value of $0.03314$ for the amplitude given to Qiskit.
The experimental data and the analytical fits are shown in Fig.~\ref{fig-rz}.
It is evident from the figure that the truncation choice is plausible, since the mean absolute error improves approximately by a factor of 5 for the model formula ($4.15 \times 10^{-3}$) compared to the Lorentzian ($20.68 \times 10^{-3}$).
%\footnote{The Lorentzian \st{baseline} was inspired from~\cite{ibm-textbook} where it is used for fitting Gaussian model resonance peaks.} 
Also, it can be noticed from the plot that the tails follow the Rosen-Zener
expression almost perfectly, which is reflected in the results. 
Neither of the two fits is overfit as their OFI are smaller than the 
threshold by about 2 orders of magnitude.
The corresponding values for the fitting parameters $\epsilon_0$ and $\epsilon_1$ are $0.033$ and $0.921$.

The standard deviation in the resonant frequency was found to be $11.39$~kHz
for the analytical fit versus $42.96$~kHz for the Lorentzian. Similar to the resonance in the Rabi model, we find that the analytical fit improves the accuracy almost by a factor of 4.

%%%%%%%%%%%%%%%%%%%%%%%%%%%%%%%%%%%%%%%%%%%%%%%%%%%%%%%%%%%%%%%%%%%%%%%%%%%%%%%%%%%%%%%%%%%%%%%
\subsection{Exponential pulse: Demkov model\label{sect-demkov}}
\subsubsection{Transition probability}

The Demkov model~\cite{Demkov1963} assumes an exponentially varying Rabi frequency, 
\begin{equation}
    \label{eq-demkov-def}
    \Omega(t) = \Omega_0 e^{-|t-T/2|/\tau}, 
\end{equation}
and a constant detuning $\Delta$.
The transition probability for this model is given by the exact expression~\cite{Vitanov1992,Vitanov1993} 
    \begin{align}
        \P01^{(0)}(\Delta) = &\left(\frac{\pi \omega}{2}\right)^2 \sech^2{\left(\frac{\pi% 
        \delta}{2} \right)} \notag \\
        \times &\left\{ \text{Re}{\left[J_{\left(1+i\delta\right) / 2} \left(\frac{\omega}{2}\right) J_{-\left(1+i\delta\right)/2}\left(\frac{\omega}{2}\right)\right]}\right\}^2,
        \label{eq-demkov}
    \end{align}
where $J_\nu(z)$ is the Bessel function of the first kind.
In this formula $\delta = \Delta \tau$ and $\omega = \Omega_0 \tau$. 

Alternatively, one can use the Rosen-Zener conjecture~\cite{Rosen1932} to determine the transition line profile $\P01^{(0)}(\Delta)$ of the Demkov model
\be
    \label{eq-demkov-pre}
    \P01^{(0)}(\Delta) = \sin^2{\left(S/2\right)} \left|\frac{1}%
    {\Omega_0 \Tilde{\tau}}\int_{-\infty}^{\infty}{\Omega(t)e^{i\Delta t}\,dt}%
    \right|^2,
\ee
where $S$ is the pulse area and $\Tilde{\tau} = \int_{-\infty}^{\infty} \Omega(t)\,dt$, which becomes $\Tilde{\tau} = 2 \tau$ in the case of an exponential pulse. 
In the original Rosen-Zener paper~\cite{Rosen1932}, the pulse area $S$ is given by $S = \Omega_0 \Tilde{\tau}$, but a later modification~\cite{Robiscoe1983, Conover2011} uses 
\be
    \label{eq-conover}
    S(\Delta) = \Tilde{\tau} \sqrt{\Omega_0^2 + a \Delta^2}, 
\ee
where $a$ is a free parameter that depends on the pulse shape. 
It is $0$ for the Rosen-Zener model and $1$ for the Rabi model.
Assuming that $\tau > 0$ and $\Delta$ are real, Eq.~\eqref{eq-demkov-pre} and Eq.~\eqref{eq-conover} can be used to derive an expression for the Demkov model transition probability
\be
    \label{eq-demkov-rzc}
    \P01^{(0)}(\Delta) = \frac{\sin^2{\left(\tau \sqrt{\Omega_0^2 + a \Delta^2}\right)}}{1 + \Delta^2 \tau^2}.
\ee
The free parameter $a$ is fixed to the value $a = 0.158$ by fitting Eq.~\eqref{eq-demkov-rzc} to the numerical solution of the Demkov model with the same parameters as in the experiment. 
%The value $a = 0.158$ is used in the demonstration.
Similar to the other models, Eqs.~\eqref{eq-demkov} and \eqref{eq-demkov-rzc} are preprocessed to account for effects caused by experimental setup limitations using the formula in Eq.~\eqref{eq-readout}.

\subsubsection{Implementation}
\label{sect-exp-demkov}

\begin{figure}[tb]
    \centering
    \includegraphics[width=\linewidth]%
    {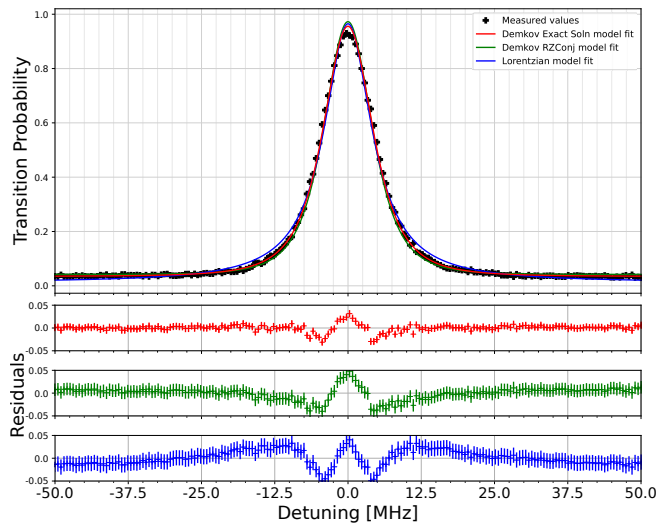}
    \caption{(Colour online) The transition line profile for the exponential Demkov model: measured data (black crosses), fit using the analytic formula with the Bessel functions Eq.~\eqref{eq-demkov} (red), the formula derived from the Rosen-Zener conjecture Eq.~\eqref{eq-demkov-rzc} (green) and a Lorentzian curve (blue) for comparison.}
    \label{fig-demkov}
\end{figure}%

The implementation on the ibm\_perth system used an exponential pulse of the form shown in Eq.~\eqref{eq-demkov-def}. 
The pulse duration was $T\approx 294.73$~ns and the pulse width was $\tau\approx 21.33$~ns. 
The amplitude of the Rabi frequency was $\Omega_0\approx 11.80$~MHz, 
corresponding to an amplitude of $0.05182$ in the Qiskit framework. 
The readings and the analytical fits are shown in Fig.~\ref{fig-demkov}.
We see that the approach to the demonstration is also very similar to 
the one with the Rosen-Zener model. 
The profile tails are described accurately by the exact formula and the Rosen-Zener conjecture~(RZC); however, some inconsistencies emerge in the near-resonance zone with all three fitting models.
The MAE for the first analytic fit (Eq.~\eqref{eq-demkov}) was $4.72 \times 10^{-3}$, while for the RZC one (Eq.~\eqref{eq-demkov-rzc}) it was $9.59 \times 10^{-3}$, compared to $13.40 \times 10^{-3}$ for the Lorentzian. 
The MAE is once again better for the analytical models, although here we observe an improvement by a factor of 3, compared to 4 in the Rabi and Rosen-Zener models.

The MAE value for the Rosen-Zener conjecture fit is larger than the exact model formula fit by roughly a factor of 2.

The values used for the $\epsilon_0$ and $\epsilon_1$ parameters are
$0.035$ and $0.921$ for the exact formula fit and
$0.043$ and $0.930$ for the Rosen-Zener conjecture formula.

Overall, 
the uncertainty of the resonance frequency was decreased by a factor of $2.2$ using 
the analytic fit: 
from $31.70$~kHz to $14.70$~kHz with the Bessel function formula and $24.71$~kHz with the RZC formula. The behaviour is expected and all three fits pass the 
overfitting criterion. 

%%%%%%%%%%%%%%%%%%%%%%%%%%%%%%%%%%%%%%%%%%%%%%%%%%%%%%%%%%%%%%%%%%%%%%%%%%%%%%%%%%%%%%%
\subsection{Gaussian pulse}
\subsubsection{Transition probability}

So far we have only considered exact analytic solutions for the models. 
For the Gaussian model we have the Rabi frequency
\begin{equation}
    \label{eq-gauss-def}
    \Omega(t) = \Omega_0 \exp{\left(-\frac{\left(t - {T}/{2}%
    \right)^2}{\tau^2}\right)}%
\end{equation}
and we use an approximate analytical model for the transition probability~\cite{Vasilev2004}.
The model can be solved using the Davis-Pechukas approach~\cite{Vasilev2004} yielding the following expression for the transition probability %$\P01$
\begin{equation}
    \label{eq-gauss}
    \P01^{(0)}(\Delta) = \frac{\sin^2{\left[\operatorname{Re}{\left(\mathcal{D}(\Delta;%
    \tau_0^+)\right)}\right]}} {\cosh^2{\left[\operatorname{Im}{\left(\mathcal{D}%
    (\Delta;\tau_0^+)\right)}\right]}},
\end{equation}
where 
\be
\mathcal{D}(\Delta; \tau_0^+) = \Delta \tau \int_0^{\tau_0^+}{\sqrt{\alpha^2 %
e^{-2\tau^2}+1}}\,d\tau,
\ee 
is the Davis-Pechukas integral for the first transition point $\tau_0^+$, which is 
the first (complex-valued, in the upper half-plane) zero of the quasienergy splitting 
$\varepsilon(t)=\sqrt{\Omega^2(t) + \Delta^2}$. 
The resulting values for the real part of the Davis-Pechukas integral are
\begin{align}
    \nonumber&\text{Re}\left(\mathcal{D}(\Delta;\tau_0^+)\right) \approx \Delta \tau \Bigg\{\left(\sqrt{\alpha^2+1}\right)-1)\Bigg. \\
    \nonumber&\times \sqrt{\frac{1}{2} \ln{\left(\frac{\alpha^2}{((1+\nu(\sqrt{\alpha^2+1}-1))^2-1}\right)}} \\
    & \Bigg.+\frac{1}{2}\sqrt{\sqrt{\left(L^2+\pi^2\right)}+L}\Bigg\},
\end{align}
where $$L = \ln{\left(\frac{\alpha^2}{\mu(2-\mu)}\right)},$$ $\alpha = \Omega_0 / \Delta$, $\mu\approx 0.316193$ and $\nu\approx %
0.462350$ are constants. 
The imaginary part of the Davis-Pechukas integral is
\be
    \operatorname{Im}{\left(\mathcal{D}(\Delta;\tau_0^+)\right)} \approx %
    \frac{1}{2}\Delta \tau\sqrt{\sqrt{4 \ln^2{\left(m\alpha\right)}%
    +\pi^2}-2\ln{\left(m\alpha\right)}},
\ee
where $m\approx 1.311468$ is a constant~\cite{Vasilev2004}.

Another approach to solving the Gaussian model utilises the Rosen-Zener conjecture~\cite{Rosen1932} (Eq.~\eqref{eq-demkov-pre}) and the subsequent corrections to the pulse area~\cite{Robiscoe1983, Conover2011} (Eq.~\eqref{eq-conover}), similar to the Demkov model.
Using Eq.~\eqref{eq-demkov-pre} and Eq.~\eqref{eq-conover}, and assuming that $\tau>0$ and $\Delta$ are real, we find an approximation for the transition line profile of the Gaussian model of the form
\be
    \P01^{(0)}(\Delta) = \sin^2{\left(\sqrt{\frac{\pi}{2}} \tau \sqrt{\Omega_0^2 + a \Delta^2}\right)} \exp{\left(-\Delta^2 \tau^2\right)}.
    \label{eq-gauss-rzconj}
\ee
Fitting Eq.~\eqref{eq-gauss-rzconj} to the numerical solution of the Schr\"odinger equation of the Gaussian model, we fix the value $a = 0.676$ and apply the preprocessing step, according to Eq.~\eqref{eq-readout}.

\subsubsection{Implementation}\label{sect-exp-gauss}

\begin{figure}[tb]
    \centering
    \includegraphics[width=\linewidth]{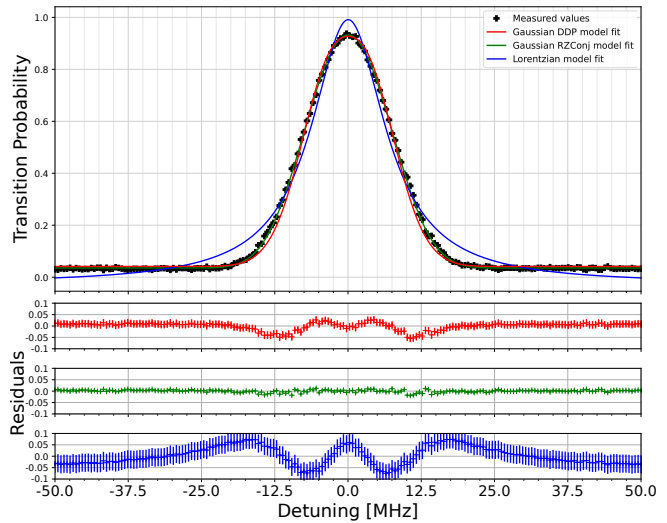}
    \caption{(Colour online) The transition line profile for the Gaussian model: 
    measured data (black crosses),
    fit using the Gaussian model DDP formula in Eq.~\eqref{eq-gauss} 
    (red), the Gaussian RZC model (green), and a Lorentzian curve (blue) for comparison.}
    \label{fig-gauss}
\end{figure}

For this analysis a pulse with duration $T\approx 112.14$~ns and width $\tau\approx21.33$~ns was used. 
The magnitude of the Rabi frequency amplitude was $\Omega_0\approx 13.25$~MHz. 
The equivalent Qiskit value for the amplitude fed to the ibm\_perth system was $0.05730$.
The measured data and the analytical fits are shown in Fig.~\ref{fig-gauss}. 
The results suggest that both approximate expressions \eqref{eq-gauss} and \eqref{eq-gauss-rzconj} are in better agreement with the demonstrated results than the Lorentzian.
The mean absolute errors are $12.27 \times 10^{-3}$ for the DDP-approximated model and $4.21 \times 10^{-3}$ for the Rosen-Zener conjecture versus $33.43 \times 10^{-3}$ for the Lorentzian, which signals a reduction of the MAE by a factor of 2.5 using the DDP, and 8 using the RZC model. 
As is evident from the MAE indices, the RZC model performs better than the DDP approximation, reducing the MAE by a factor of $3$.
These results confirm the relevancy of the approximate solutions~\cite{Vasilev2004}. 
The values of the $\epsilon_0$ and $\epsilon_1$ parameters are $0.042$ and $0.885$ for the DDP fit and $0.035$ and $0.897$ for the RZC fit.

The standard deviation of the resonance frequency was $35.68$~kHz for the analytic DDP model, $13.54$~kHz for the analytic RZC model, compared to $87.92$~kHz for the Lorentzian.
This constitutes approximately a decrease of the uncertainty by a factor of 2.5 with the DDP model and 6.5 with the RZC model. 
The OFI indices of the three fits are also well within the overfitting threshold.

%%%%%%%%%%%%%%%%%%%%%%%%%%%%%%%%%%%%%%%%%%%%%%%%%%%%%%%%%%%%%%%%%%%%%%%%%%%%%%%%%%%%%%%
\subsection{Squared hyperbolic-secant pulse: the sech$^2$ model}
\subsubsection{Transition probability}

The sech$^2$ model is another model which has no exact analytic solution. The Rabi frequency takes the shape
\be
    \label{eq-sech2-def}
    \Omega(t) = \Omega_0 \sech^2{\left(\frac{t - {T}/{2}}{\tau}\right)},
\ee
where the pulse is centred at $T/2$ and the detuning $\Delta$ is constant.
The analytical form of the transition probability was derived using the modified Rosen-Zener conjecture (Eq.~\eqref{eq-demkov-pre}) \cite{Rosen1932, Robiscoe1983}. 

Using Eqs.~\eqref{eq-demkov-pre} and~\eqref{eq-sech2-def} and assuming that $\tau>0$ and $\Delta$ are real, we solve the integral and find that
\be
    \label{eq-sech2}
    \P01^{(0)}(\Delta) = \frac
    {\left(\pi \Delta \tau\right)^2 \sin^2(\frac{1}{2}S(\Delta))} 
    {\sinh^2\left(\frac{1}{2}\pi\tau\Delta\right)}.
\ee
Similar to the Demkov model Rosen-Zener conjecture derivation, instead of the usual $S=\Omega_0 \Tilde{\tau}$, we use a modification of the form $S(\Delta)=\Tilde{\tau}\sqrt{\Omega_0^2 + a \Delta^2}$, as shown in~\cite{Robiscoe1983, Conover2011}. 
Again, $a$ is a free parameter that is fixed at $a = 0.449$ and measurement errors are accounted for by carrying out preprocessing as in Eq.~\eqref{eq-readout}. 

\begin{figure}[tb]
    \centering
    \includegraphics[width=\linewidth]%
    {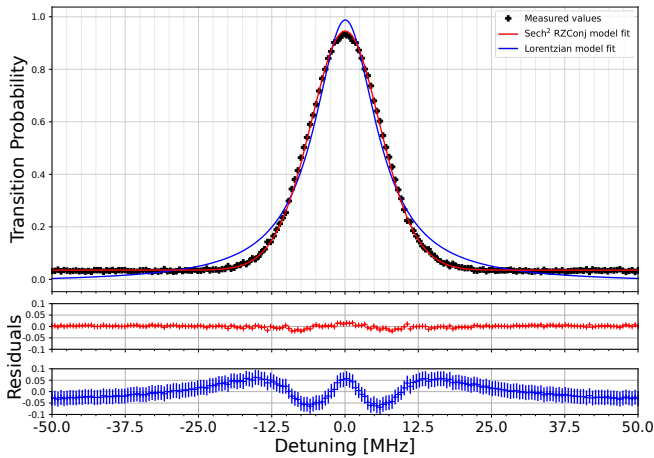}
    \caption{(Colour online) The transition line profile for the sech$^2$ model: 
    measured data (black crosses),
    fit using the sech$^2$ model formula (Eq.~\eqref{eq-sech2}) (red) and a 
    Lorentzian curve (blue) for comparison.}
    \label{fig-sech2}
\end{figure}%

\subsubsection{Implementation}\label{sect-exp-sech2}

In this demonstration, the sech$^2$ model was implemented on the ibm\_perth quantum computer using a discretised form of the pulse shape, shown in Eq.~\eqref{eq-sech2-def}. 
The duration of the pulse (truncation point) was $T \approx 176.93$~ns and the pulse width was $\tau \approx 21.33$~ns. The value for the Rabi frequency amplitude was $\Omega_0 \approx 11.76$~MHz. %35460561.388$~rad/s. 
This is equivalent to an amplitude of $0.05124$ in the Qiskit package.

The experimental data and the analytical fits are shown in Fig.~\ref{fig-sech2}. 
The analytical formula matches the measured values far better than the Lorentzian fit, resulting in a smaller value for the mean absolute error: 
$4.47 \times 10^{-3}$ for the formula versus $27.55 \times 10^{-3}$ for the Lorentzian fit, or more than a 6-fold improvement. 
%However, a certain inconsistency can be noticed in the width of the analytical profile, compared to the experiment. 
The values for the $\epsilon_0$ and $\epsilon_1$ fitting parameters for the sech$^2$ model fit are $0.035$ and $0.910$ respectively.

A strong improvement in the resonance frequency was achieved, as the standard deviation of the resonant value was decreased by a factor of 5 --- from $67.72$~kHz to $13.12$~kHz. 
No overfitting was detected in either of the fits.

% Rabi
% Model SI: 0.39785307006403703
% Baseline SI: 1.5451212777860175
% 0.014115949021476736 +- 0.001896157617755425
% 0.014242625122675238 +- 0.007255763637401145
% deviation: -0.00012667610119850188 +- 0.0074994346235702745
% Rosen-Zener
% Model SI: 0.6861643416805209
% Baseline SI: 3.6366916696391995
% 0.05448057198155315 +- 0.011370782086489065
% 0.05169723349391511 +- 0.044950398470389744
% deviation: 0.0027833384876380374 +- 0.046366291720443174
% Gaussian DDP
% Gaussian RZ Conjecture
% Model 1 SI: 0.8863730337712872
% Model 2 SI: 0.4602640845367976
% Baseline SI: 4.021729327257195
% 0.033263268298766205 +- 0.010734710354950995
% 0.03193522080585191 +- 0.006066008834445641
% 0.027961224374961215 +- 0.04386120652190849
% deviation1: 0.00530204392380499 +- 0.04515572437645309
% deviation2: 0.003973996430890699 +- 0.04427868449646038% Demkov
% Model SI: 0.9602957947977007
% Baseline SI: 1.5019229173050734
% 0.011490295754751507 +- 0.011659346456859112
% 0.010957301470943039 +- 0.01759061321609242
% deviation: 0.0005329942838084682 +- 0.02110379191328512
%Sech$^2$
% Model SI: 0.646704438098349
% Baseline SI: 4.029387297660105
% 0.025463338565288617 +- 0.009869881473579674
% 0.019672614931419127 +- 0.045640839005498125
% deviation: 0.005790723633869489 +- 0.04669583220618635%

%%%%%%%%%%%%%%%%%%%%%%%%%%%%%%%%%%%%%%%%%%%%%%%%%%%%%%%%%%%%%%%%%%%%%%%%%%%%%%%%%%%%%%%
\subsection{Discussion}

The values of the mean absolute error (MAE) and the standard deviation of the resonance frequency (SDRF)
 for all models are summarized in Table \ref{table-di}.
For all models the reduction of both MAE and SDRF is by a factor of 4 to 8. 
The only exception is the exponential Demkov model for which the improvement is about 180\% due to the nonanalyticity of the pulse shape at its maximum, which makes the Lorentzian fit quite reasonable. 
Yet, in absolute terms (i.e. disregarding the Lorentzian fits), the analytic fit for the Demkov model is nearly as good as those for the other analytic models, delivering comparable values of the MAE.

% \begin{table}[t!]
% \begin{tabular}{|l|r|r|r|r|}
% \hline
% Model        & Analytic & \st{Baseline} \hl{Lorentzian} & Analytic & \st{Baseline} \hl{Lorentzian} \\ 
%  & \st{DI} \hl{MAE} & \st{DI} \hl{MAE} & SDRF (kHz) & SDRF (kHz) \\
% \hline
% Rabi           & 0.40 & 1.55 & 1.9  & 7.3 \\ \hline
% Rosen-Zener    & 0.69 & 3.64 & 11.4 & 45.0 \\ \hline
% Demkov Bessel  & 0.80 & 1.50 & 10.1 & 17.6 \\ \hline
% Demkov RZC     & 0.79 & 1.50 & 9.4  & 17.6 \\ \hline
% Gauss. DDP     & 0.89 & 4.02 & 10.7 & 43.9 \\ \hline
% Gauss. RZC     & 0.46 & 4.02 & 6.1  & 43.9 \\ \hline
% Sech$^2$       & 0.65 & 4.03 & 9.9  & 45.6 \\ \hline
% \end{tabular}
% \caption{Values for the \st{divergence index (DI)} \hl{mean absolute error (MAE)} and the standard deviation of the resonance frequency (SDRF)  for each of the models, 
% including the value for the \st{baseline} \hl{Lorentzian} fit.}
% \label{table-di}
% \end{table}

\begin{table}[t!]
\begin{tabular}{|l|r|r|r|r|}
\hline
Model        & Analytic & Lorentzian & Analytic & Lorentzian \\ 
 & MAE & MAE & SDRF & SDRF  \\
 & ($\times 10^{-3}$) & ($\times 10^{-3}$) & (kHz) & (kHz) \\
\hline
Rabi           & 9.41  & 36.21 & 79.8  & 315.2 \\ \hline
Rosen-Zener    & 4.15  & 20.68 & 11.39 & 42.96 \\ \hline
Demkov Bessel  & 4.72  & 13.40 & 14.70 & 31.70 \\ \hline
Demkov RZC     & 9.59  & 13.40 & 24.71 & 31.70 \\ \hline
Gauss. DDP     & 12.27 & 33.43 & 35.68 & 87.92 \\ \hline
Gauss. RZC     & 4.21  & 33.43 & 13.54 & 87.92 \\ \hline
Sech$^2$       & 4.47  & 27.55 & 13.12 & 67.72 \\ \hline
\end{tabular}
\caption{Values for the mean absolute error (MAE) and the standard deviation of the resonance frequency (SDRF) for each of the models, 
including the value for the Lorentzian fit.}
\label{table-di}
\end{table}

%%%%%%%%%%%%%%%%%%%%%%%%%%%%%%%%%%%%%%%%%%%%%%%%%%%%%%%%%%%%%%%%%%%%%%%%%%%%%%%%%%%%%%%
\section{Conclusion}

We have presented experimental results of the measurement of the qubit transition line profiles for 
five different pulse shapes of the driving microwave field obtained using one of the online IBM 
Quantum processors. 
We have compared these experimental profiles with two types of fits: one using the exact or approximate analytic solution for the respective pulse shape and the other using the routinely exploited Lorentzian function.
We have shown that the measured data show much better resemblance to the respective analytic solutions than to intuitive fitting functions, according to the mean absolute errors~(see Table~\ref{table-di}). 
Nevertheless, we have outlined some  minor inconsistencies due to the unavoidable truncation of the driving pulse amplitude. 
For all but the exponential model the MAE index shows an improvement by a factor of 4 to 8.
For the exponential model the improvement is by 180\% only, because the Lorentzian behaves much better than for the sech, sech$^2$ and Gaussian models.

Such a comparison provides quantitative measures of both the accuracy of the respective theoretical model, the quality of the driving field, and the ``quantumness'' of the qubit regarding its coherence properties.

We point out that for three of the pulse shapes --- Gaussian, exponential and sech$^2$ --- we have used analytic description based on the 90-year old Rosen-Zener conjecture \cite{Rosen1932}, augmented by a modification based on earlier work by Robiscoe \cite{Robiscoe1983} and Conover \cite{Conover2011}. 
It is known that this augmented conjecture delivers the exact results for the rectangular and sech pulse shapes.
Much to our surprise, the rather simple formulas obtained by it for the Gaussian, exponential and sech$^2$ pulse shapes prove very accurate.
Therefore we can claim that the present work presents an experimental demonstration of the validity of this conjecture for these pulse shapes.

Apart from demonstrating fine features of qubit dynamics, the results in this paper can have an useful practical application: more accurate determination of the exact resonance, which can result in improved fidelity of the quantum gates. 
For the five pulse shapes studied here, we have found differences in the resonant frequency, as determined by the analytic formulae and the Lorentzian fits, in the range of a few dozens of kHz. 
The superiority of the new values for the resonant frequency has been verified by a reduction of its standard deviation by a factor of about $4$ in four of the models; the only exception to this was 
the Demkov model, where the uncertainty was reduced by a factor of $2.2$ only.
This difference may affect the fidelity of the quantum gates, especially for longer pulse durations.
Therefore, the present paper provides an easy solution to one of the experimental issues in the quest for ultrahigh fidelity of gate operations.

Finally, although the results in this paper have been demonstrated on a specific quantum computing platform --- the transmon-based IBM Quantum system --- the conclusions should be generally valid on any other physical platform.

\acknowledgements

This research is supported by the Bulgarian national plan for recovery and resilience, contract BG-RRP-2.004-0008-C01 (SUMMIT), project number 3.1.4. We acknowledge the use of IBM Quantum services for this work. The views expressed are those of the authors, and do not reflect the official policy or position of IBM or the IBM Quantum team.

%%%%%%%%%%%%%%%%%%%%%%%%%%%%%%%%%%%%%%%%%%%%%%%%%%%%%%%%%%%%%%%%%%%%%%%%%%%%%%%%%%%%%%%

\appendix

\section{Two-state transmon physics}
\label{sect-transmon}

Any purely two-level quantum system is a qubit (e.g. the electron spin). 
However, in practice some multi-level systems are also used as two-level systems. 
The main condition is that there is sufficient anharmonicity present, which allows the transitions between the first two levels to be controlled independently from the next possible transition. 
In other words, the anharmonicity is the difference between the energy of the $0\rightarrow 1$
and the $1\rightarrow 2$ transitions~\cite{Krantz2019}. 
One system which satisfies the "qubit requirements" is the transmon, which is formulated from \textit{transmission-line shunted plasma oscillation qubit}~\cite{Koch2007}. 

In order to explain its physics, it is easier to start with a normal LC circuit and derive the Hamiltonian for it, before we jump into the transmon. 
The Hamiltonian for a normal LC circuit is equivalent to the one for a linear harmonic oscillator 
if we express the mass $m$ for the capacitance $C$, the momentum $p$ for the electrical charge $Q$ and the resonant frequency $\omega_r = 1/\sqrt{LC}$. 
Therefore, the usual Hamiltonian for the linear harmonic oscillator reads
\be
    \mathcal{H}_{LHO} = \frac{p^2}{2m} + \frac{m \omega^2 x^2}{2}
\ee
becomes 
\be
    \label{eq-LHO}
    \mathcal{H}_{LC} = \frac{Q^2}{2C} + \frac{\Phi^2}{2L} = \frac{1}{2} CV^2 + \frac{1}{2} LI^2
\ee
where $\Phi$ is the magnetic flux, $I$ is the electric current, $L$ is the
inductance and $V$ is the voltage~\cite{Krantz2019}. 

We can also introduce non-linear versions of the charge and the flux - the reduced charge~$n = Q/{(2e)}$, where $e$ is the electron charge, and the reduced flux~$\phi = 2\pi \Phi/{\Phi_0}$, with $\Phi_0 = h/(2e)$ being the magnetic flux quantum~\cite{Krantz2019}. 
Then we substitute the linear charge and flux in Eq.~\eqref{eq-LHO} and yield 
\be
    \label{eq-transmon1}
    \mathcal{H}_{\text{LC}} = \frac{4e n^2}{2C} + \frac{\Phi^2_0\phi^2}{8\pi^2L}.
\ee
The constants in front of $n$ and $\phi$ can be simplified by introducing 
\begin{align}
    \nonumber &E_C=\frac{e^2}{2C} \text{ and}\\ &E_L = \frac{\Phi^2_0}{4\pi^2 L},
\end{align}
where $E_C$ is the charging energy needed to add another Cooper pair to the island, and $E_L$ is the inductive energy. 
Then we obtain an equation for the Hamiltonian in terms of these constants which reads
\be
    \label{eq-transmon2}
    \mathcal{H}_{\text{LC}} = 4E_C n^2 + \frac{1}{2} E_L\phi^2.
\ee

To some extent, the transmon is similar to the plain LC circuit, but instead of the inductance we have a so-called Josephson junction~\cite{Josephson1962, Josephson1964}. 
This is a non-linear circuit element that does not allow any dissipation, and essentially
acts as a non-linear version of the inductance. 
To find the transmon Hamiltonian $H_\text{T}$, we need to incorporate its properties into the equation. 
Its properties are best described by the identities for $I$ and $V$
\be
I = I_c \sin{(\phi)}, \qquad
    V = \frac{\hbar}{2e} \frac{\d\phi}{\d t}.
\ee
These expressions yield 
\be
    \mathcal{H}_{\text{T}} = 4 E_\Sigma n^2 + E_J \cos{\phi}
    \label{eq-transmon3}
\ee
for the transmon Hamiltonian, with
\be
   E_\Sigma=\frac{e^2}{2C_\Sigma}, \qquad 
   E_J = \frac{I_c \Phi_0}{2\pi},
\ee
where $C_\Sigma = C_s + C_J$ is formed by both the shunt capacitance $C_s$ and the self-capacitance $C_J$ of the Josephson junction. 
The second term in Eq.~\eqref{eq-transmon3} is nonlinear but it can be Taylor expanded for small values of $\phi$. 
This results in 
\be
    \label{eq-taylor}
    E_J \cos{(\phi)} = E_J - \frac12 E_J \phi^2 + \frac{1}{24} E_J \phi^4 - %
    \mathcal{O}(\phi^6),
\ee
where we retain only the first three terms. 
The second term is quadratic and if we limit ourselves there, we would be looking at a 
quantum harmonic oscillator~(QHO). 
However, the third term is quartic in $\phi$ and it is responsible for the uneven shift in energy levels and the resulting differences in excitation energies between them~(the anharmonicity). 
It is also clear from the sign of the third term that the anharmonicity $\alpha = E_{1\rightarrow 2} - E_{0\rightarrow 1}$ is negative and the transition energy decreases for the higher levels. 

The energy contributions by the charge and the flux parts are determined by the ratio $E_J/E_\Sigma$.  Experimentally, it was found that flux noise was much easier to control than charge noise. If $E_J \ll E_\Sigma$ the qubit became much more sensitive to charge noise, so this was ruled out. Instead, a suitable value for the ratio was found to be $E_J/E_\Sigma \gtrsim 50$~\cite{Nakamura1999, Vion2002, You2003, Duty2004}. 
The methods to achieving such a ratio mainly aim at reducing $E_\Sigma$ by having a capacitor with high capacitance, effectively decreasing charge noise. 
That is the case in the transmon qubit~\cite{Koch2007}. 
 
By taking the three leading terms in the Taylor expansion of Eq.~\eqref{eq-taylor}, we can express the Hamiltonian in its second quantisation form as 
\be
    H_{\text{T}} = E_0 a^\dag a + \frac{\alpha}{2} a^\dag a^\dag a a,
\ee
where $|\alpha| \ll E_0$ which is consistent with an anharmonic oscillator. 
If $|\alpha|$ is large enough and gates are constructed so as to prevent leakage (as in the so-called DRAG gates~\cite{Werninghouse2021}), excitation to higher levels is suppressed
and we can effectively treat the system as a two-level quantum system. 
The Hamiltonian can be reduced to 
\be
H_{\text{T}} = E_0 \frac{\sigma_z}{2},
\ee
where $\sigma_z = \operatorname{diag}{(1,-1)}$ is the corresponding Pauli matrix. 
As long as the $E_J/E_\Sigma \gtrsim 50$ condition holds, we have a functional superconducting qubit.

% \bibliography{pulse_shape} 
% \begin{comment}

% \end{comment}

\end{document}